\documentclass[useAMS,usenatbib]{mn2e}
\usepackage{rotating}
\usepackage{journals}
\usepackage{graphicx, subfigure}
\usepackage{xspace}
\usepackage{amssymb}

\def\farcs{\hbox{$.\!\!^{\prime\prime}$}}
\def\degr{\hbox{$^\circ$}}

\def\kms{$\rmn{km\,s^{-1}}$}

\def\ang{$\rmn{\AA}$\xspace}

\def\chan{{\it Chandra}\xspace}
\def\swift{{\it Swift}\xspace}

\def\integral{{\it INTEGRAL}\xspace}

\def\src{IGR~J19308+0530\xspace}

\newcommand\msun {$M_{\odot}$\xspace}

\newcommand{\mdot}{$\dot{m}$}

\newcommand{\ergsec}{$\rmn{erg\,s^{-1}}$}

\LARGE \normalsize \title[The mass ratio in \src]{\src: Roche lobe overflow on to a compact object from a donor 1.8 times as massive \thanks{Observations made with the HERMES spectrograph at the Mercator Telescope, operated on La Palma by the Flemish Community, at the Spanish Observatorio del Roque de los Muchachos of the Instituto de Astrof'sica de Canarias. HERMES is supported by the Fund for Scientific Research of Flanders, the Research Council of K.U.Leuven and the Fonds National Recherches Scientific (FNRS), Belgium, the Royal Observatory of Belgium, the Observatoire de Gen\'eve, Switzerland and the ThŸringer Landessternwarte Tautenburg, Germany.}}

\author[Ratti et al.]  {E.M.~Ratti$^{1}$\thanks{email : e.m.ratti@sron.nl}, T.F.J.~van~Grunsven$^{1,2}$, M.A.P.~Torres$^{1,3}$, P.G.~Jonker$^{1,2,3}$, \newauthor J. C. A.~Miller-Jones$^{4}$,  J. W.T.~Hessels$^{5,7}$, H.~Van Winckel$^6$, M.~van~der~Sluys$^{2,8}$, \newauthor G.~Nelemans$^{2,6}$ \\
$^1$SRON, Netherlands Institute for Space Research, Sorbonnelaan 2, 3584~CA, Utrecht, The Netherlands\\ 
$^2$Department of Astrophysics/IMAPP, Radboud University Nijmegen, Heyendaalseweg 135,6525 AJ, Nijmegen, The Netherlands \\
$^3$Harvard--Smithsonian Center for Astrophysics, 60 Garden Street, Cambridge, MA~02138, U.S.A.\\
$^4$International Centre for Radio Astronomy Research - Curtin University, GPO Box U1987, Perth, WA 6845, Australia \\
$^5$ASTRON, the Netherlands Institute for Radio Astronomy, Postbus 2, 7990 AA, Dwingeloo, The Netherlands \\ 
$^6$Instituut voor Sterrenkunde, K.U.Leuven, Celestijnenlaan 200D, B-3001 Leuven, Belgium \\
$^7$Astronomical Institute "Anton Pannekoek," University of Amsterdam, Science Park 904, 1098 XH Amsterdam, The Netherlands \\
$^8$Nikhef National Institute for Subatomic Physics, Science Park 105, 1098 XG Amsterdam, The Netherlands 
}

\begin{document}

\maketitle

\begin{abstract}  
We present phase-resolved spectroscopy and photometry of the optical counterpart to the X--ray binary \src.
Ellipsoidal modulations in the lightcurve show that the F-type companion star in the system is Roche-lobe filling. The
optical spectra are dominated by absorption features from the donor star, with $\sim$10-20\% disc contribution to the
optical continuum. We measure an orbital period of 14.662$\pm$0.001 hours,  a radial velocity semi-amplitude for the
companion star of K$_2=$91.4$\pm$1.4\,\kms and a rotational broadening of $v\sin i=108.9\pm0.6$\,\kms. From K$_2$ and
$v\sin i$, given that the donor star is filling its Roche lobe, we derive a mass ratio of $q= $M$_2/$M$_1=$
1.78$\pm0.04$, which is typically considered to be too large for stable Roche-lobe overflow. Our observations support
an inclination of $\sim$50 degrees. The accretor in \src is most likely a white dwarf, although a neutron star cannot
entirely be excluded. 

\end{abstract}


\section{introduction}

Intermediate-mass X--ray binaries (IMXBs) are binary systems where a compact object - black hole, neutron star (NS) or
white dwarf (WD) - is accreting matter from a companion star of spectral type A or F. IMXBs are rarely observed (see,
e.g., the catalogue from \citealt{Liu07} and the 2012 version of the RK catalogue
\citealt{Rit03}\footnote{http://physics.open.ac.uk/RKcat/}). The majority of accreting WDs in binaries, in fact, belongs to the class of cataclysmic variables (CVs), which have late type secondaries ($M_2\lesssim M_1$) and  $P\lesssim 6$ hours
\citep{Kni11-1}. NSs or BHs, instead, are typically observed in X--ray binaries (XRBs) hosting either a massive O-B donor star
(high-mass XRBs, HMXBs) driving accretion via stellar wind, or a late M or K dwarf secondary star (low-mass XRBs, LMXBs) accreting
via Roche lobe overflow. The reason for the observed rarity of IMXBs especially among NS and WD systems is that, when the companion is more massive than
the accretor but not massive enough to have strong winds, wind accretion proceeds at a very low rate and Roche-lobe accretion is thought to be unstable. For NSs and WDs in IMXBs mass flows
from the more massive to the lighter star and angular momentum conservation shrinks the orbit, leading to enhanced
mass transfer. The bright X-ray binary phase is therefore intense and short-lived, causing an observational bias towards LMXBs,
CVs and HMXBs \citep{Tau06}.  Nevertheless, IMXBs could be a large fraction of the XRB population and have an
important role in understanding their evolution \citep{Pod01}. Cyg~X--2 and Her~X--1 are though to have started as
IMXBs, even though the measured mass ratio is currently $<1$. 
\newline 
\src was discovered by \integral \citep{Bir06} and observed by \swift \citep{Rod08}. An association with the
star TYC\,486-295-1, classified as an F8 star in the survey by \citet{McC49}, was made using
the \swift position \citep{Rod08}. This was confirmed using an accurate \chan position of the X-ray source
\citep{Rat10}. Considering typical parameters of an F8 star, \citet{Rod08} suggested \src to be a L/IMXB in quiescence
or a CV at a distance of $\lesssim1$ $\rmn{kpc}$. 
\newline 
Here, we present phase-resolved optical spectroscopy and photometry of \src, in order to measure the orbital
period $P$,  the radial velocity semi-amplitude $K_2$ and the projected rotational velocity $v\sin i$ of the companion
star, and the system inclination $i$. In a Roche lobe filling system, $K_2$ and $v\sin i$ allow us to infer the ratio
$q=$M$_2/$M$_1$ between the mass of the secondary and the primary star in the system (\citealt{Wad88}, see also
\citealt{Gra92}) and, knowing P and $i$, to solve the system mass function.

\section{Observations and data reduction}
\label{sec:data}


In total twenty-two high-resolution spectra of \src were collected. Observations were made on two nights in 2010
Mar., one night in 2010 Apr.~and 9 nights in 2010 Jun.~using the fiber spectrograph High Efficiency and Resolution Mercator Echelle
Spectrograph (HERMES), mounted at the Mercator telescope in La Palma \citep{Ras11}. The typical exposure time was 1200
seconds. The spectra have a dispersion of 0.027 \ang/pixel at 5000 \ang and cover the
wavelength range 3770-7230 \ang. The fiber aperture is 2\farcs5 on the sky, but the presence of a slicer mimics a narrow slit providing a resolution of $\sim$85000 irrespective of the seeing. The template star HD185395, of spectral type F4\,{\sc V} (later found to be the closest match to the spectral type of the target, see below) was observed with the same
settings for 360 seconds on 2009 Aug.~5. The extraction of the spectra was performed through the dedicated automated
data reduction pipeline HermesDRS. For each spectrum we selected two regions for the analysis, one covering the
H$\gamma$ and H$\beta$ lines (4280-5250 \ang) and one around the H$\alpha$ line (5950-6700 \ang), which we will refer
to as S1 and S2, respectively. These regions were selected as they are rich in stellar lines, with little contamination
from interstellar features. A good fit of the continuum was achieved in each region with a polynomial function of
order 9. We normalized the spectra dividing by~the~polynomial~fit. 


We also performed time-resolved photometry of \src, with the 80-cm IAC80 telescope at the Observatorio del Teide in
Tenerife equipped with the CAMELOT CCD imager. The observations were obtained during part of three nights between 2010
Jul.~30 and Aug.~02, by cycling through the Sloan {\it g}$^\prime$, {\it  r}$^\prime$, {\it i}$^\prime$ and {\it
z}$^\prime$-band filters. Three consecutive exposures were obtained in each filter, with integration times ranging
from 6 to 24 sec depending on the filter and seeing conditions.  Standard stars were not taken due to non-photometric
weather. After debiasing and flat-fielding the images using dome flat-field observations (with standard routines in {\sc IRAf}) the instrumental magnitudes
of \src and four comparison stars were computed by means of aperture photometry. Differential lightcurves were then
obtained for \src with respect to the comparison star TYC 486-968-1.  We also extracted lightcurves for TYC 486-968-1
using  the other  comparison stars. No significant variability was detected in the $i^\prime$-band (r.m.s.$ \lesssim
0.006$ mag), whereas for the other bands a larger scatter was observed due to weather conditions (clouds and Calima). The maximum
departure from the mean value was of 0.1 mag. Given that the ellipsoidal modulation in \src is small in amplitude we
decided to model only the $i^\prime$-band lightcurves.  

\section{Analysis and results}
\label{sec:res}

\begin{figure}
\includegraphics[ width=8.5cm, angle=0]{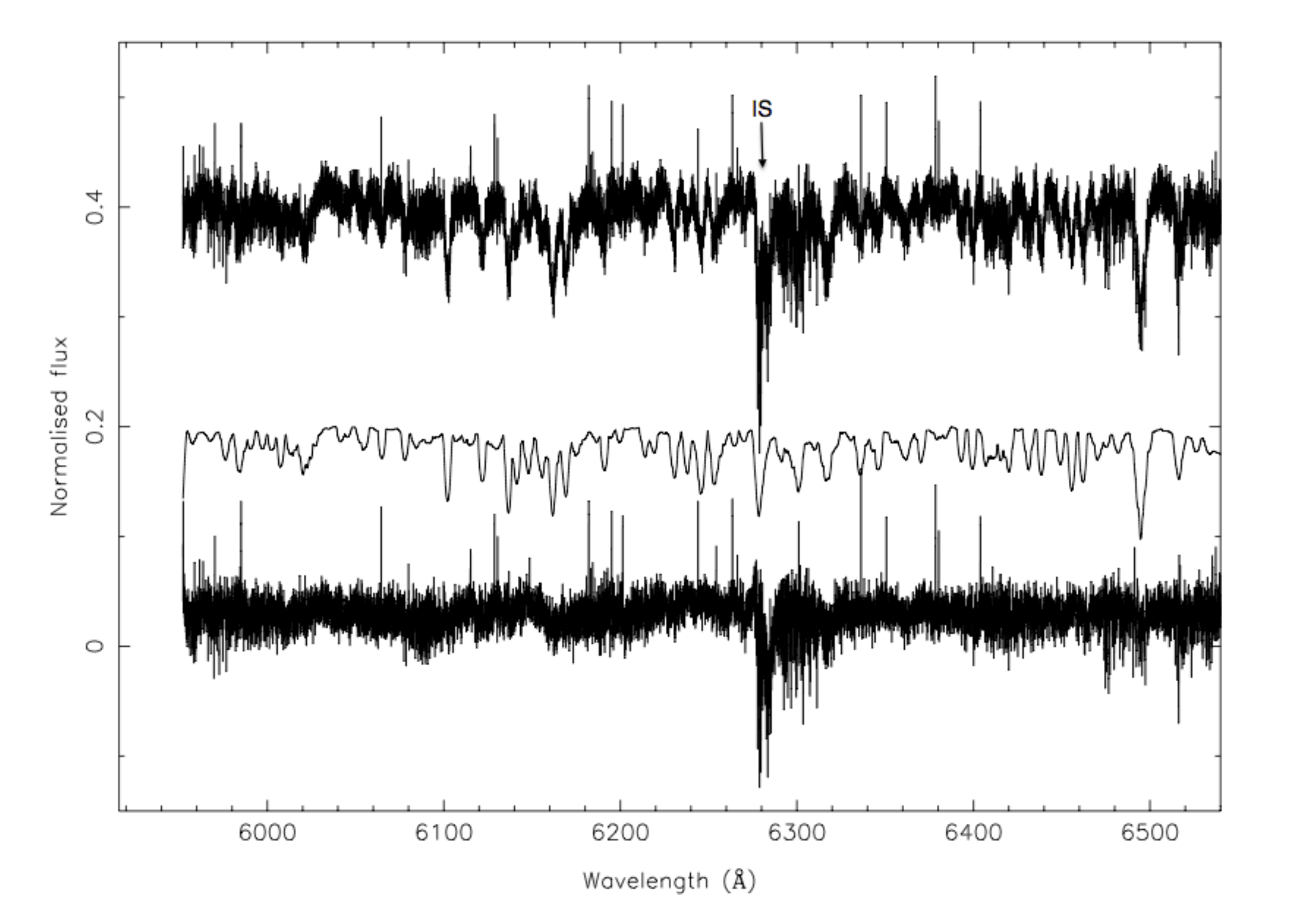} \\
\vskip -0.3cm
\caption{\label{fig:optsub} Optimal subtraction of the spectrum of the F4V stellar template HD185395 from the spectrum of \src. The image shows the part of spectral region S2 used to measure $v \sin i$. The interstellar feature at $\sim$6280 \ang was masked. The template and target spectrum are offset by 0.4 and 0.2 in y. 
}
\end{figure}

\subsection{Spectroscopy}
\label{sec:spec}
\begin{figure}
\includegraphics[width=8cm, angle=0]{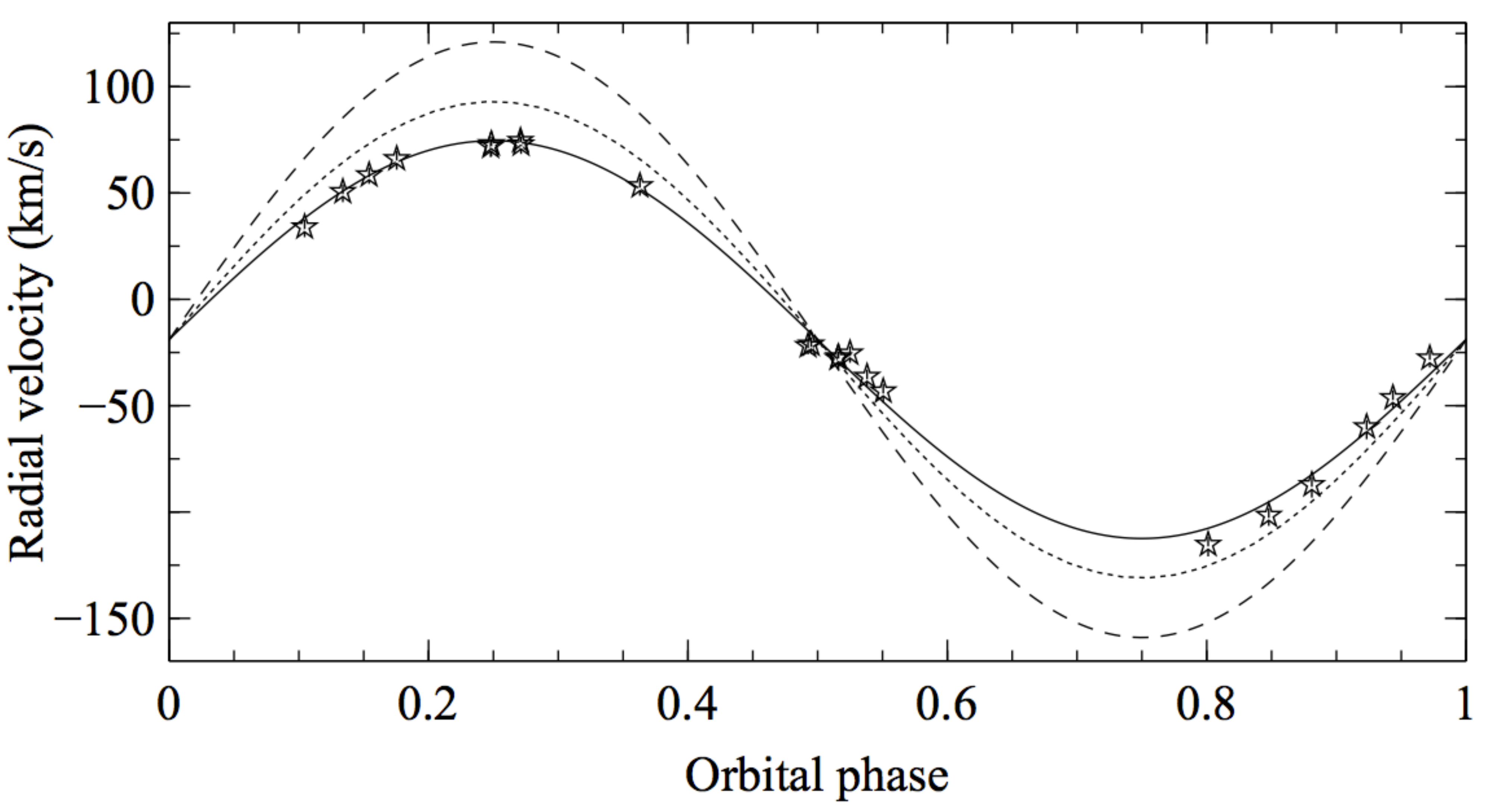} \\
\vskip -0.4cm
\caption{\label{fig:rvc} Radial velocity curve from the absorption lines in the \src spectra (region S1) and its best fitting sinusoid (solid line). For comparison we plot circular orbits with $K_2=$112\,\kms (dotted line) and $K_2=$140\,\kms (dashed line), providing $q=1.4$ and $q=1$ respectively for the observed $v\sin i$.}
\end{figure}

The spectra are dominated by absorption features from the secondary star (\ref{fig:opt sub}). Although no emission line is directly
visible, residual emission in the Balmer lines appears when subtracting the spectra one from another, after correcting
for the orbital shift of the lines. The emission component is variable in intensity and wavelength and slightly shifted with respect to the absorption line. \\
\indent Using the package {\sc molly} developed by T.Marsh, we measured the orbital velocity of the companion star in \src by cross-correlating the spectra of the target with that of the
template star. As the absorption lines are rotationally broadened in the target spectrum, a broadening of 100 \kms was
applied to the template (see below on rotational broadening), improving the cross-correlation. The Balmer lines and
interstellar features were masked. \\ We performed a fit of the velocities versus time with a sine function, with
$K_2$, the systemic radial velocity $\gamma$, $T_0$, and $P$ as free parameters. $T_0$ was constrained to be near the
middle of the time span over which the observations were taken, and such that phase 0 is at the inferior conjunction
of the companion star. The best-fitting sinusoid provided a $\chi^2$ of 57.6 for the region S1, and 41.76 for S2 (19
d.o.f.). In both cases, the uncertainties on the parameters were estimated assuming that the sinusoidal model was
correct, and we scaled the errors on the velocities to reach a reduced $\chi^2$ of $\sim$1. The values of $P$,
$T_0$ and $\gamma$ measured in S1 and S2 are consistent at the 1$\sigma$ level, $K_2$ is consistent at the 2 $\sigma$
level. The error-weighted average of the parameters gives $P=0.61092\pm0.00003$ days, $T_0=2455330.8169\pm0.0023$
HJD/UTC, K$_2=91.4\pm1.4~\rmn{km\,s^{-1}}$ and $\gamma=-18.5\pm0.9~\rmn{km\,s^{-1}}$. Figure \ref{fig:rvc} shows the
radial velocity curve (rvc) folded on the above period.  The value of $\gamma$ is in the reference frame of the
template star used for the cross-correlation, whose systemic radial velocity is $-28\pm0.9$~\kms \citep{Wil53}. The
systemic radial velocity of \src is therefore $-46.5\pm1.2$~\kms. 

We obtained a set of high signal-to-noise template spectra from the UVES Paranal Observatory Project (UVESPOP,
\citealt{Bag03}) of A, F and early G stars of luminosity class {\sc V}, {\sc III} and {\sc IV}. We subtracted each
template spectrum from the Doppler-corrected average of the \src spectra between orbital phase 0.9 and 0.1. The reason
for choosing this range in phase is that the oblate shape of the Roche-lobe filling companion star and the possible
presence of irradiation from the compact object could cause asymmetries in the line profiles, which are minimised close
to phase 0. We performed a $\chi^2$ test on the residuals of the subtraction: the template resulting in the minimum
$\chi^2$ provides our best  estimate for the source spectral type. In particular, we adopted the optimal subtraction
procedure implemented in {\sc molly}, where the templates are multiplied by a factor 0$<$\,{\it f}\,$<1$ before the
subtraction, representing the fractional contribution of light from the secondary star (1 minus the disc veiling). The
factor $f$ is found by minimizing the difference between the residuals and a smoothed version of itself.  Before
doing the subtraction, the UVESPOP spectra were shifted to the rest frame of the average target spectrum, and degraded
to match the sampling~and~line~broadening~of~the~latter. 
\newline 
The procedure favors an F4-F6\,{\sc V} companion star, with a disc veiling of $10-20$\%. The same spectral type and
veiling are obtained when considering an average of the target spectra around phase 0.5, suggesting little irradiation
on the inner face of the companion star. 

To measure $v\sin i$, we compared the spectrum of the template star with the Doppler-corrected average of
the \src spectra between phase 0.9 and 0.1.   The observed full-width at half-maximum (FWHM) of the absorption lines
in the target spectra is determined by the intrinsic line width (expected to be dominated by $v\sin i$), broadened by
the instrumental resolution profile and smeared by the motion of the  companion star during the integration time of
one observation.   In order to account for the smearing, we made as many copies of the template spectrum as the number
of target spectra we used for the average and we artificially smeared each copy of the template by  $2 \pi T K
\cos(2\pi\phi)/P$, where $T$ is the duration of one exposure on \src and $\phi$ the phase of one of the \src spectra
we averaged. After that, we averaged the smeared template and broadened the resulting spectrum with different values
of  $v\sin i$. For each $v\sin i$, we performed an optimal subtraction of the broadened template from the averaged
spectrum of \src: again the broadening which gives the minimum  $\chi^2$ provides a measure of the actual $v\sin i$
(Figure \ref{fig:optsub}). A value of 0.5 was assumed for the limb darkening. We masked interstellar features and the lines from the Balmer series. \\ In order to estimate the uncertainty on
$v\sin i$, we included this procedure in a Monte Carlo simulation, following \citet{Stee07}. We copied each target
spectrum 500 times, using a bootstrap technique where the input spectrum is resampled by randomly selecting data
points from it. The bootstrapping maintains the total number of data points in the spectrum. For each bootstrap copy,
one value of $v\sin i$ is measured as described above. The distribution of $v\sin i$ obtained from the 500 copies is
well described by a Gaussian, whose mean and r.m.s.~provides the best-fit $v\sin i$ and its 1$\sigma$
error.  As template and target spectra are acquired with the same instrument, the instrumental resolution profile is
not affecting our~measurement.  \newline The weighted average of the results from S1 and S2, consistent at the
1$\sigma$ level, is $v\sin i=108.9\pm0.6$\,\kms. With $v\sin i$ and $K_2$, we calculated the system mass ratio
$q=M_2/M_1$  from the relation ${v \sin i \over K_2}=(1+q){0.49q^{2/3}\over 0.6q^{2/3}+\ln(1+q^{1/3})}$
\citep{Hor86_b}, obtaining $q=$1.78$\pm$0.04. 

\begin{figure}
\vskip -0.7cm
\hskip -0.8cm
\includegraphics[width=9.5cm, angle=0]{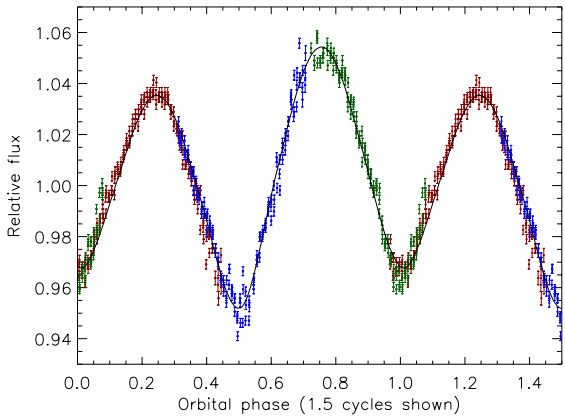}
\vskip -0.7cm
\caption{\label{lice} $i^\prime$-band CAMELOT lightcurve, folded using the ephemeris from the radial velocity data. The colours indicate the three observing nights (Jul. 30 in red, Jul. 31 in green and Aug. 1 in blue). The drawn line shows a fit with ellipsoidal modulations plus a disc with a bright disc spot.}
\end{figure}
\vskip -1.3cm

\subsection{Ellipsoidal modulation and system inclination} 

Figure \ref{lice} shows the $i^\prime$-band lightcurve for \src obtained by phase folding the CAMELOT data on the
ephemeris determined in Section \ref{sec:spec}. The lightcurve displays the typical signature of ellipsoidal
variation, with two unequal minima, but in addition asymmetric maxima (\citealt{O'C51} and \citealt{Wils09}).
\newline We modeled the $i^\prime$-band lightcurve using the XRbinary program written by E.L. Robinson.
A reasonable fit to the data (reduced $\chi^2\sim7$, 403 d.o.f.) is obtained with a model assuming an F4{\sc V} secondary star and including 30\%
disc contribution to the total light plus a disc hot-spot at phase 0.75. The disc veiling in the model is larger than observed in the spectra,
but variability is possible as the photometry was performed one month after the last spectrum was acquired. Similar
models with different assumptions about the disc properties (disc radius, height and temperature profile) also
give reasonable fits. As we have no indications to single out one preferred set of parameters, we do not provide a
formal uncertainty on $i$, but an indicative value of  $\sim$52\,\degr as a guide.  
\vspace{-0.35cm}
\subsection{Upper limits in the radio waveband}
We searched for a radio counterpart to \src with a Karl G. Jansky Very Large Array observation at 4.6 and 7.9
GHz (120 MHz bandwidth) (proposal ID 10B-238), taken on 2010 Aug. 19 with the array in its most compact
D-configuration. The on-source time was 33 minutes and the data were reduced according to the standard procedures
within the Common Astronomy Software Application \citep{McM07} software package, using the calibrator
3C 286 to set the amplitude scale and J1922+1530 to calibrate the amplitude and phase gains for the target source. IGR
J19308+0530 was not detected at either frequency.  A 3-sigma upper limit to the source flux of 54 $\mu$Jy/beam was derived \\
\indent To search for radio pulsations from a potential pulsar in this system, we used the Westerbork Synthesis Radio
Telescope and the PuMa2 backend \citep{kss08}.  We observed for 1 hr from both 310$-$380 MHz (Obs. ID 11205182, 29
Aug. 2012) and from 1300$-$1460 MHz (Obs. ID 11205210, 31 Aug. 2012).  The predicted dispersion measure (DM) for a 300
pc distance along this line-of-sight (see Discussion) is only $\sim 4$pc $\rmn{cm^{-3}}$ \citep{cl02}.  For each data
set, we used the PRESTO software suite \citep{ran01} to search a set of trial DMs up to 30 pc $\rmn{cm^{-3}}$.  
No obvious radio pulsar signal was detected after an acceleration search. From these observations, we can place conservative flux density limits of $S_{350} < 0.9$
mJy and $S_{1400} < 0.4$ mJy for any pulsar present in the system, assuming it is beamed towards us.

\section{Discussion and conclusion}
\label{sec:disc}

We performed a dynamical study of the system \src through optical spectroscopy and photometry. The optical spectra are
dominated by the companion star, with no evidence of irradiation, no emission features visible from the accretion flow
besides a partial filling in of the Balmer lines and $10-20$\% disc contribution to the continuum. The secondary star
is most likely of spectral type F4-6\,{\sc V}. Ellipsoidal modulations are detected on the $\sim$14.6 hour orbit. From
phase-resolved spectroscopy we measure an extreme value for the binary mass ratio of $q=$1.78$\pm$0.04. The lightcurve modeling provides a reasonable fit to the data with a disc+hot-spot model at
$i\sim52$\degr. Solving the mass function $f(M_2):  M_1{sin^3i\over (1+q)^2} ={P{K_2}^3\over2\pi G}=$0.03\,\msun with
this inclination, we obtain the following indicative masses:  M$_1\sim$0.8\,\msun and
M$_2\sim$1.4\,\msun. The masses are consistent with a WD accretor and an F4{\sc V} donor (typical mass of
$\sim$1.37\,\msun, Mamajek's list 2011\footnote{http://www.pas.rochester.edu/~$\sim$emamajek/ \\
EEM\_dwarf\_UBVIJHK\_colors\_Teff.dat}). If the inclination is lower, which we cannot exclude based on these data, the
masses will increase, allowing a scenario with a NS primary if $i\lesssim45$\degr. However, in this case the companion
star would be over-massive for the spectral type, which is unusual for XRB and CVs. 


Assuming an F4\,{\sc V} mass donor with a radius equal to that of the Roche lobe for our best estimated masses, the magnitude of the companion is $M_V=2.7$ in the visual band (half a magnitude brighter
than for a typical F4\,{\sc V} star, Mamajek 2011). Comparing with the apparent magnitude of \src ($m_V=10.95\pm0.13$, converted from $m_{VT}$ in the Tycho catalogue), we estimate a distance range of $300-450\,\rmn{pc}$, for an extinction between 
$N_H=(0-2.6)\times10^{21}$\,cm$^{-2}$ (\citealt{Dic90}, where $N_H$ is converted into $A_V$ following \citealt{Guv09}).
\newline Combining the systemic radial velocity with the source proper motion reported in the UCAC3 catalogue \citep{Zac10}, we
computed the Galactic space velocity components of \src using the method of
\citet{Joh87}.  Assuming that the Local Standard of Rest (LSR) participates in the Galactic rotation at 254\,\kms
\citep{Rei09} and a distance of $375\pm75\,\rmn{pc}$, the derived peculiar velocity is 45.3$\pm$2.9\,\kms. This rules out a large asymmetric kick from a supernova. \\
\indent The X--ray luminosity of \src measured by \chan in 2007 \citep{Rat10} is $5\times10^{29}-4\times10^{30}$\ergsec, for distance in the above range of $300-450\rmn{pc}$ and the corresponding $N_H$. 
In the same way, UV \swift observations provide $4\times10^{30}-1\times10^{31}$$\rmn{erg\,s^{-1}\,\AA^{-1}}$ at 2500 \ang. 
A scenario where \src is not in full contact and the X--rays are from coronal activity of the companion seems unlikely. First, coronal activity usually does not produce prolonged high energy emission, while the source was discovered from a stack of \integral observations. Second, coronal activity could explain the observed X--ray luminosity only for the earliest spectral type allowed by the observations combined with very little $N_H$, as for F stars the ratio between the X--rays and bolometric flux is $log(L_X/L_{bol})\lesssim-4.6$ for $v\sin i \sim100$ \citep{Wal83}. Finally, the spectra indicate disc contribution to the continuum, and the emission detected in the Balmer lines shows a slight, variable velocity offset with respect to the radial velocity curve of the companion star which is not expected in case they originate from coronal activity. The emission line component also seems weak for an highly active star. 
 A hot WD of $\sim$60000$\,\rmn{K}$ alone could account for the UV emission, but not for $L_X$ unless thermonuclear burning is happening on its surface. The few intermediate-mass CVs with long orbital period that are known do appear as super soft sources (SSSs), showing soft X--ray spectra possibly due to stable hydrogen burning on the WD  (e.g.
\citealt{Kah06}). However, with a luminosity of 10$^{36}-10^{38}$ \ergsec, SSSs are much more luminous than what we observe from \src. 
As the wind mass loss expected for the secondary spectral type is low
($\sim10^{-14}$\msun\,yr$^{-1}$, \citealt{Cra11}) wind accretion can not account for the observed X--ray
luminosity. \\ We conclude that \src is most likely in contact, consistent with the
ellipsoidal modulations observed in the lightcurve. This makes the source particularly  interesting among the scarcely
populated class of IMXBs since it shows Roche lobe overflow at a low accretion rate \mdot. This is unusual, as Roche lobe
accretion  with an intermediate mass companion and such a large mass ratio is typically considered unstable, with an
intense and short-lived accretion phase (e.g., \citealt{Tau06}). Donor stars of mass around 1.4\,\msun have thin or
non-existent convective envelopes \citep{ver81} which implies that the instability does not proceed on a  dynamical
timescale. While a convective envelope star would react to mass loss by expanding, a star with a mostly radiative
envelope will shrink, slowing down the mass transfer. However, with a mass ratio of 1.8 the radius of the Roche lobe
will reduce faster in response to mass transfer than the radius of the companion, triggering a thermal instability
\citep{Tau06}.  
Using the binary stellar-evolution code originally
developed by Eggleton (\citealt{Yak05} and references therein), we modeled evolutionary tracks that allow for periods of stable Roche-lobe accretion onto a WD with masses and orbital periods consistent with our findings, although at higher \mdot\xspace values than implied by the $L_X$ (see also the appendix in
\citealt{Pod03}). The low $L_X$ can be explained if we are observing a short lived phase of low \mdot, or if the mass transfer is non-conservative (a fully non-conservative scenario requires a mass loss rate from the system of $10^{-10}-10^{-9}$\msun/yr). There is currently no direct evidence in favour of or against the presence of mass loss from the system. \\
\indent From an observational point of view, an overestimate of $q$ could be due to uncertainties on the limb darkening. However, even assuming a limb darkening of 0 (instead of 0.5) reduces $v\sin i$ by only a few \kms, still providing a high $q$ of $\sim1.6$. 
\vspace{-0.5cm}
\section*{Acknowledgments} \noindent  PGJ and GN acknowledge support from a  VIDI grant from the Netherlands
Organisation for Scientific Research. We thank T. Marsh for {\sc Molly}, E.L. Robinson for his XRbinary code and F. Verbunt for useful discussion. The National Radio Astronomy  Observatory is a facility of the National Science Foundation operated 
under cooperative agreement by Associated Universities, Inc.

\vskip -1.5cm
\bibliographystyle{mn_new} \bibliography{F8.bib}

\begin{thebibliography}{33}
\expandafter\ifx\csname natexlab\endcsname\relax\def\natexlab#1{#1}\fi

\bibitem[{{Bagnulo} et~al.(2003){Bagnulo}, {Jehin}, {Ledoux}, {Cabanac},
  {Melo}, {Gilmozzi}, \& {ESO Paranal Science Operations Team}}]{Bag03}
{Bagnulo}, S., {Jehin}, E., {Ledoux}, C., {Cabanac}, R., {Melo}, C.,
  {Gilmozzi}, R., {ESO Paranal Science Operations Team}, 2003, The Messenger,
  114, 10

\bibitem[{{Bird} et~al.(2006)}]{Bir06}
{Bird}, A.~J., et~al., 2006, \apj, 636, 765

\bibitem[{{Cordes} \& {Lazio}(2002)}]{cl02}
{Cordes}, J.~M., {Lazio}, T.~J.~W., 2002, ArXiv Astrophysics e-prints

\bibitem[{{Cranmer} \& {Saar}(2011)}]{Cra11}
{Cranmer}, S.~R., {Saar}, S.~H., 2011, \apj, 741, 54

\bibitem[{{Dickey} \& {Lockman}(1990)}]{Dic90}
{Dickey}, J.~M., {Lockman}, F.~J., 1990, \araa, 28, 215

\bibitem[{{Gray}(1992)}]{Gra92}
{Gray}, D.~F., 1992, {The observation and analysis of stellar photospheres.},
  Camb. Astrophys. Ser., Vol. 20, C.U.P

\bibitem[{{G{\"u}ver} \& {{\"O}zel}(2009)}]{Guv09}
{G{\"u}ver}, T., {{\"O}zel}, F., 2009, \mnras, 400, 2050

\bibitem[{{Horne} et~al.(1986){Horne}, {Wade}, \& {Szkody}}]{Hor86_b}
{Horne}, K., {Wade}, R.~A., {Szkody}, P., 1986, \mnras, 219, 791

\bibitem[{{Johnson} \& {Soderblom}(1987)}]{Joh87}
{Johnson}, D.~R.~H., {Soderblom}, D.~R., 1987, \aj, 93, 864

\bibitem[{{Kahabka}(2002)}]{Kah02}
{Kahabka}, P., 2002, ArXiv Astrophysics e-prints

\bibitem[{{Kahabka}(2006)}]{Kah06}
{Kahabka}, P., 2006, Advances in Space Research, 38, 2836

\bibitem[{{Karuppusamy} et~al.(2008){Karuppusamy}, {Stappers}, \& {van
  Straten}}]{kss08}
{Karuppusamy}, R., {Stappers}, B., {van Straten}, W., 2008, \pasp, 120, 191

\bibitem[{{Knigge} et~al.(2011){Knigge}, {Baraffe}, \& {Patterson}}]{Kni11-1}
{Knigge}, C., {Baraffe}, I., {Patterson}, J., 2011, \apjs, 194, 28

\bibitem[{{Liu} et~al.(2007){Liu}, {van Paradijs}, \& {van den Heuvel}}]{Liu07}
{Liu}, Q.~Z., {van Paradijs}, J., {van den Heuvel}, E.~P.~J., 2007, \aap, 469,
  807

\bibitem[{{McCuskey}(1949)}]{McC49}
{McCuskey}, S.~W., 1949, \apj, 109, 426

\bibitem[{{McMullin} et~al.(2007){McMullin}, {Waters}, {Schiebel}, {Young}, \&
  {Golap}}]{McM07}
{McMullin}, J.~P., {Waters}, B., {Schiebel}, D., {Young}, W., {Golap}, K.,
  2007, in {R.~A.~Shaw, F.~Hill, \& D.~J.~Bell}, ed., Astronomical Data
  Analysis Software and Systems XVI, vol. 376 of \emph{Astronomical Society of
  the Pacific Conference Series}, p. 127

\bibitem[{{O'Connell}(1951)}]{O'C51}
{O'Connell}, D.~J.~K., 1951, Publications of the Riverview College Observatory,
  2, 85

\bibitem[{{Podsiadlowski} et~al.(2001){Podsiadlowski}, {Rappaport}, \&
  {Pfahl}}]{Pod01}
{Podsiadlowski}, P., {Rappaport}, S., {Pfahl}, E., 2001, The influence of
  binaries on stellar population studies, Dordrecht: Kluwer Academic
  Publishers, 2001, xix, 582 p.~Astrophysics and space science library (ASSL),
  Vol.~264.~ISBN 0792371046, p.355, 264, 355

\bibitem[{{Podsiadlowski} et~al.(2003){Podsiadlowski}, {Han}, \&
  {Rappaport}}]{Pod03}
{Podsiadlowski}, P., {Han}, Z., {Rappaport}, S., 2003, \mnras, 340, 1214

\bibitem[{{Ransom}(2001)}]{ran01}
{Ransom}, S.~M., 2001, {New search techniques for binary pulsars}, Ph.D.
  thesis, Harvard University

\bibitem[{{Raskin} et~al.(2011)}]{Ras11}
{Raskin}, G., et~al., 2011, \aap, 526, A69

\bibitem[{{Ratti} et~al.(2010){Ratti}, {Bassa}, {Torres}, {Kuiper},
  {Miller-Jones}, \& {Jonker}}]{Rat10}
{Ratti}, E.~M., {Bassa}, C.~G., {Torres}, M.~A.~P., {Kuiper}, L.,
  {Miller-Jones}, J.~C.~A., {Jonker}, P.~G., 2010, \mnras, 408, 1866

\bibitem[{{Reid} et~al.(2009){Reid}, {Menten}, {Brunthaler}, {Zheng},
  {Moscadelli}, \& {Xu}}]{Rei09}
{Reid}, M.~J., {Menten}, K.~M., {Brunthaler}, A., {Zheng}, X.~W., {Moscadelli},
  L., {Xu}, Y., 2009, \apj, 693, 397

\bibitem[{{Ritter} \& {Kolb}(2003)}]{Rit03}
{Ritter}, H., {Kolb}, U., 2003, \aap, 404, 301

\bibitem[{{Rodriguez} et~al.(2008){Rodriguez}, {Tomsick}, \& {Chaty}}]{Rod08}
{Rodriguez}, J., {Tomsick}, J.~A., {Chaty}, S., 2008, \aap, 482, 731

\bibitem[{{Steeghs} \& {Jonker}(2007)}]{Stee07}
{Steeghs}, D., {Jonker}, P.~G., 2007, \apjl, 669, L85

\bibitem[{{Tauris} \& {van den Heuvel}(2006)}]{Tau06}
{Tauris}, T.~M., {van den Heuvel}, E.~P.~J., 2006,
  {Formation~and~evolution~of~compact~stellar~X-ray~sources}, C.U.P., p. 623

\bibitem[{{Verbunt} \& {Zwaan}(1981)}]{ver81}
{Verbunt}, F., {Zwaan}, C., 1981, \aap, 100, L7

\bibitem[{{Wade} \& {Horne}(1988)}]{Wad88}
{Wade}, R.~A., {Horne}, K., 1988, \apj, 324, 411

\bibitem[{{Walter}(1983)}]{Wal83}
{Walter}, F.~M., 1983, \apj, 274, 794

\bibitem[{{Wilsey} \& {Beaky}(2009)}]{Wils09}
{Wilsey}, N.~J., {Beaky}, M.~M., 2009, Society for Astronomical Sciences Annual
  Symposium, 28, 107

\bibitem[{{Wilson}(1953)}]{Wil53}
{Wilson}, R.~E., 1953, Carnegie Institute Washington D.C.~Publication, 0

\bibitem[{{Yakut} \& {Eggleton}(2005)}]{Yak05}
{Yakut}, K., {Eggleton}, P.~P., 2005, \apj, 629, 1055

\end{thebibliography}

\end{document}